\documentclass[letterpaper,12pt]{article}

\usepackage{graphics,graphicx}

\def\eps@scaling{1.0}%
\newcommand\plotone[1]{%
 \typeout{Plotone included the file #1}
 \centering
 \leavevmode
 \includegraphics[width={\eps@scaling\columnwidth}]{#1}%
}%
\newcommand\plottwo[2]{{%
 \typeout{Plottwo included the files #1 #2}
 \centering
 \leavevmode
 \columnwidth=.45\columnwidth
 \includegraphics[width={\eps@scaling\columnwidth}]{#1}%
 \hfil
 \includegraphics[width={\eps@scaling\columnwidth}]{#2}%
}}%

\makeatletter
\renewcommand{\section}{\@startsection%
{section}{1}{0mm}{-\baselineskip}%
{0.5\baselineskip}{\normalfont\Large\bfseries}}%
\makeatother

\begin{document}
\pagestyle{plain}
\pagenumbering{arabic}

\begin{center} 
\bfseries{
\Large{The Behavior of Matter Under Extreme Conditions}\\
\vskip 0,3in
\large{A White Paper Submitted to the {\it Astro2010} Decadal Survey of Astronomy and 
Astrophysics}
}
\\
\vskip 1in
F. Paerels$^1$, 
M. M\'endez$^2$,
M. Agueros$^1$,
M. Baring$^3$,
D. Barret$^4$,\\
S. Bhattacharyya$^5$,
E. Cackett$^6$,
J. Cottam$^7$,
M. D\'iaz Trigo$^8$,
D. Fox$^9$, \\
M. Garcia$^{10}$,
E. Gotthelf$^{\ 1}$,
W. Hermsen$^{11}$,
W. Ho$^{12}$,
K. Hurley$^{13}$,
P. Jonker$^{11}$,\\
A. Juett$^7$,
P. Kaaret$^{14}$,
O. Kargaltsev$^9$,
J. Lattimer$^{15}$,
G. Matt$^{16}$,
F. \"Ozel$^{17}$,\\
G. Pavlov$^9$,
R. Rutledge$^{18}$,
R. Smith$^{10}$,
L. Stella$^{19}$,
T. Strohmayer$^7$,\\
H. Tananbaum$^{10}$,
P. Uttley$^{12}$,
M. van Kerkwijk$^{20}$,
M. Weisskopf$^{21}$,
S. Zane$^{22}$
\end{center}

\begin{center}
$^1$ Columbia University, $^2$ University of Groningen, $^3$ Rice University,\\
$^4$ CESR Toulouse, $^5$ Tata Institute of Fundamental Research, Mumbai, \\
$^6$ University of Michigan, $^7$ NASA Goddard Space Flight Center, \\
$^8$ European Space Astronomy Centre, ESA, $^9$ Penn State University, \\
$^{10}$ Harvard-Smithsonian Center for Astrophysics, $^{11}$ SRON Netherlands Institute for Space Research, $^{12}$University of Southampton, $^{13}$ University of California at Berkeley, \\ 
$^{14}$ University of Iowa, $^{15}$ SUNY Stony Brook, $^{16}$ Universit\`a degli Studi Roma Tre, \\
$^{17}$ University of Arizona, $^{18}$ McGill University,$^{19}$ Osservatorio Astronomico di Roma, \\$^{20}$ University of Toronto, $^{21}$ NASA Marshall Space Flight Center, \\
$^{22}$ Mullard Space Science Laboratory
\end{center}

\setcounter{page}{0}

\newpage


\begin{center} 
\bfseries{
\Large{The Behavior of Matter Under Extreme Conditions}
}
\\
\end{center}


\begin{abstract}
The cores of neutron stars harbor the highest matter densities known to occur in nature, up to several times the densities in atomic nuclei. Similarly, magnetic field strengths can exceed the strongest fields generated in terrestrial laboratories by ten orders of magnitude. Hyperon-dominated matter, deconfined quark matter, superfluidity, even superconductivity are predicted in neutron stars. Similarly, quantum electrodynamics predicts that in strong magnetic fields the vacuum becomes birefringent. The properties of matter under such conditions is governed by Quantum Chromodynamics (QCD) and Quantum Electrodynamics (QED), and the close study of the properties of neutron stars offers the unique opportunity to test and explore the richness of QCD and QED in a regime that is utterly beyond the reach of terrestrial experiments. Experimentally, this is almost virgin territory. 
\end{abstract}

\section{Introduction: The Fundamental Properties of Matter at High Densities}

Seven decades after the first speculation on the existence of gravitationally bound neutron configurations (Landau 1932, 1938), we still know very little about the fundamental properties of neutron stars. Initial attempts to model their mechanical properties were based on the assumption that the matter can be adequately described as a degenerate gas of free neutrons, but it has become progressively clear that the cores of neutron stars must in fact be the stage for intricate and complex collective behavior of the constituent particles. 

Over most of the range of the density/temperature phase plane, Quantum Chromodynamics (QCD) is believed to correctly describe the fundamental behavior of matter, from the subnuclear scale up. 
The ultimate constituents of matter are quarks, which are ordinarily bound in various combinations by an interaction mediated by gluons to form composite particles. At very high energies, a phase transition to a plasma of free quarks and gluons should occur, and various experiments are currently probing this low-density, high temperature limit of QCD ({\it e.g.} Tannenbaum 2006). Likewise, the QCD of bound states is beginning to be quantitatively understood; recently, the first correct calculation of the mass of the proton was announced (D\"urr et al. 2008). 

The opposite limit of high densities and low (near zero, compared to the neutron Fermi energy) temperature QCD has been predicted to exhibit very rich behavior. At densities exceeding a few times the density in atomic nuclei ($\rho \sim 3 \times 10^{14}$ g cm$^{-3}$), exotic excitations such as hyperons, or Bose condensates of pions or kaons may appear. It has also been suggested that at very high densities a phase transition to strange quark matter  may occur. When and how such transitions occur is of course determined by the correlations between the particles, and the ultra-high-density behavior of matter is governed by many-body effects. This makes the direct calculation of the properties of matter under these conditions from QCD extremely difficult.

\begin{figure}[t]
 \begin{center}
   \begin{minipage}[t]{0.5\linewidth}
     \raisebox{-8.7cm}{\includegraphics[height=3.4in]{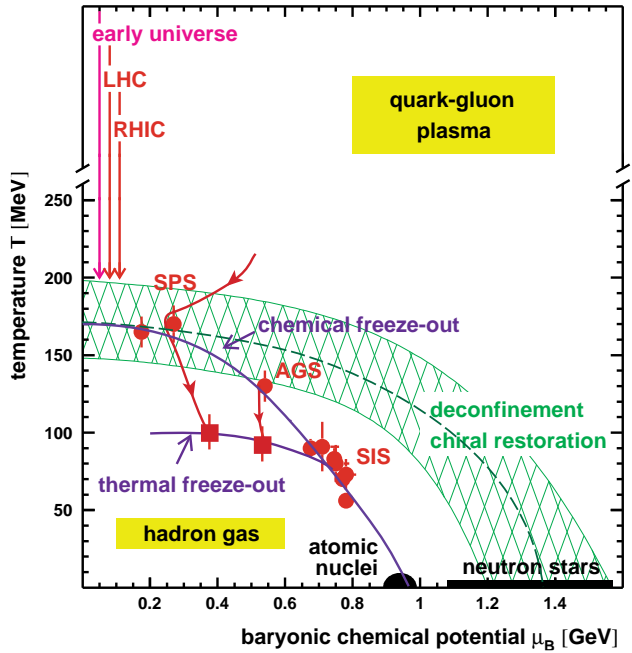}}
   \end{minipage}\hfill
   \begin{minipage}[t]{0.45\linewidth}
     \caption{
     \small{
At high energies or high densities, normal hadronic matter is expected to dissolve into a quark-gluon plasma, as shown in the upper right hand corner of the temperature-chemical potential phase plane;  the green zone marks the transition region. Terrestrial experiments probe the low-density, high-temperature limit (shown below 'Early Universe'). The high-density, low-temperature limit is expected to be marked by the appearance of exotic matter and phase transitions. This regime, beyond the density in atomic nuclei, can only be probed by astrophysical observations of neutron stars ({\it Courtesy NA49 Collaboration, SPS, CERN})
}
\label{side}}
   \end{minipage}
 \end{center}
\end{figure}

Figure 1 shows the temperature-chemical potential phase plane, in which the locus of the phase transition from the hadron gas to the quark-gluon plasma has been indicated. The only possible way to probe the high density, low temperature limit of QCD is by observations and measurements of the densest material objects in nature, neutron stars. 

\section{The Mass-Radius Relation of Neutron Stars}

The relation between pressure and density, the equation of state, is the simplest way to parameterize the bulk behavior of matter. It governs the mechanical equilibrium structure of bound stars, and, conversely, measurements of quantities such as the mass and the radius, or the mass and the moment of inertia of a star, probe the equation of state. Figure 2 shows the mass-radius plane for neutron stars, with a number of possible mass-radius relations based on various assumptions concerning the equation of state (Lattimer and Prakash 2007). Two families of solutions have been indicated: the equations of state for stars made up of bound quark states (baryons and mesons), and solutions for stars in which a phase transition has converted most of the stellar matter to strange quark matter. The former stars are gravitationally bound, and for a degenerate fermion gas, the radius generally decreases with increasing mass of the configuration. The quark stars, on the other hand, are self-bound, and exhibit, very roughly, an increase in radius with increasing mass. In the same figure, a number of constraints have been indicated, derived from, for instance, the maximum stable mass of neutron stars, and a number of broad constraints derived from observations (such as the high spin frequencies of two neutron stars). It is obvious that definitive constraints can only be derived from simultaneous measurement of masses and radii of individual neutron stars.

Effective discrimination between different families of hadronic equations of state will require a relative precision of order 10\% in mass and radius, and similar requirements apply to the strange equations of state. In order to settle the question as to whether strange stars exist in nature, the requirements depend on stellar mass. Since the hadronic and strange mass-radius relations cross in the region $\sim 1.3 - 1.8 M_{\odot}$, $12 - 16$ km (ironically, those are the textbook values for the mass and the radius!), we need to be able to probe a range of masses, or else have to rely on very difficult high-precision measurements.

\begin{figure}
\includegraphics[height=3.4in]{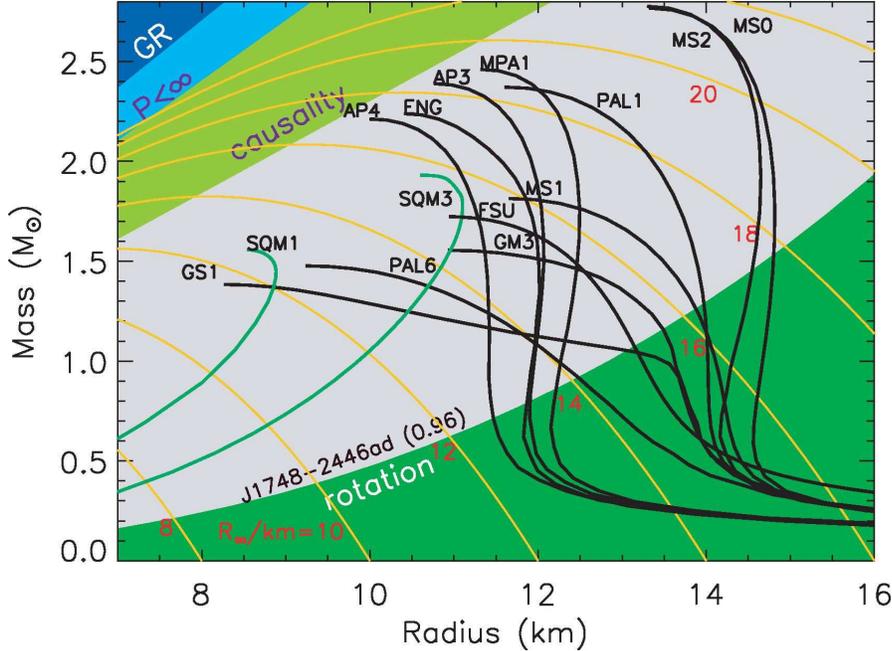}
\caption{\small{
The mass-radius relationship for neutron stars reflects the equation of state for cold superdense matter.
Mass-radius trajectories for typical EOSs are shown as black curves.  Green curves (SQM1, SQM3)
are self-bound quark stars.  Orange lines are contours of radiation
radius, $R_\infty=R/\sqrt{1-2GM/Rc^2}$.  The dark blue region is
excluded by the GR constraint $R>2GM/c^2$, the light blue region is
excluded by the finite pressure constraint $R>(9/4)GM/c^2$, and the
light green region is excluded by causality, $R>2.9GM/c^2$.  The green
region in the right-hand corner shows the region $R>R_{\rm max}$ excluded by the 716 Hz pulsar
J1748-2446ad (Hessels et al. 2006).  From Lattimer and Prakash (2007).
\label{fig2}
}
}
\end{figure}

\section{Breakthrough Potential: X-ray Observations}
\subsection{Current Status}

Neutron stars have been the subject of intensive radio observations for forty years, and this work has indeed produced a wealth of fundamental advances (see for instance, Blandford et al. 1993); probably the most famous among these is the confirmation of the prediction of the gravitational wave power emitted by a relativistic binary based on Einstein's quadrupole formula, which earned Hulse and Taylor a Nobel prize (for recent data, see for instance Weisberg and Taylor 2005). As far as the fundamental properties of the stars themselves are concerned, precise radio pulse arrival time measurements on double neutron star binaries have produced a series of exquisite mass determinations, with a weighted average stellar mass of $M_{\rm NS} = 1.413 \pm 0.028M_{\odot}$ (the error is the weighted average deviation from the mean;
see, for instance, Lattimer and Prakash 2007). 
But we need the mass and the radius, or two other quantities derived from the internal structure, simultaneously. There is currently no hope of measuring the stellar radii for the neutron stars for which we have a precise mass, from radio or other observations (with one possible exception, see below). 

Most of what we know about the fundamental properties of ordinary stars is based on a close study of the emission spectrum emerging from their photospheres. 
Neutron stars are small and relatively far away; detecting optical or UV radiation from their surfaces is extremely difficult, and in any case, the optical/UV emission will be on the Rayleigh-Jeans tail of the stellar spectrum, which is not particularly sensitive to the stellar properties. Optical radiation has been detected in a few cases (see Kaspi, Roberts, and Harding 2006 for a review), but it is likely that this corresponds to emission from an unknown fraction of the stellar surface, which makes it impossible to use these data for radius measurement. Looking for higher-luminosity objects means looking for hotter objects (assuming the radii of all neutron stars are comparable in size), and the natural wavelength band for photospheric observations is the X-ray band. 

X-ray emission originating on the surfaces of neutron stars was first detected in X-ray bursts from accreting neutron stars in Low-Mass X-ray Binaries (LMXBs), and photospheric emission has also been detected from quiescent and isolated objects ({\it e.g.} Strohmayer and Bildsten 2006; Guillot et al. 2009, and references therein). 
These data had low spectral resolution and often limited signal-to-noise; they have provided a very rough check on the order of magnitude of neutron star radii, but precise measurement awaits the development of stellar atomic spectroscopy of neutron stars, as well as a series of more exotic techniques that take advantage of general relativistic effects on the surface emission. With observations performed with the diffraction grating spectrometers on the {\it Chandra} and {\it XMM-Newton} observatories, this problem has come to the threshold of being resolved---the next step, based on sensitive, time resolved X-ray spectroscopy and energy-resolved fast photometry has the unique potential of finally providing the window into QCD that the 'Cold Equation of State' will open up.

The current observational situation is roughly the following.
There is evidence for atomic photospheric absorption in the burst spectrum of at least one accreting neutron star (EXO0748$-$646; Cottam, Paerels, \& Mendez 2002), which has led to a measurement of the gravitational redshift at the stellar surface ($z = 0.35$). Likewise, the distance to a number of hot, intermittently accreting neutron stars is known, because they are located in Globular Clusters. Once accretion ceases, the atmospheres of these stars should simply consist of pure H, and the spectrum can be calculated;
this sample of stars with known distance can be extended (Guillot et al. 2009, and references therein).
Three apparently isolated neutron stars have a parallax measurement (Walter and Lattimer 2002; Kaplan, van Kerkwijk, and Anderson 2002, 2007; Pavlov et al. 2008).

\subsection{Opportunities}

Several techniques are available with sensitive X-ray spectroscopy and fast photometry. Spectroscopic observations of X-ray bursts give the atomic absorption spectrum, which, through pressure broadening
and GR effects, is sensitive to both the acceleration of gravity at the stellar surface, as well as the redshift. Measuring two different functions of mass and radius thus gives mass and radius, separately. For neutron stars at known distance, measured fluxes compared to the flux emerging from the stellar photosphere will give the stellar radius. The continuum spectral shape is sensitive to the surface gravity, again allowing a mass and radius measurement.

\begin{figure}[t]
\centering
\includegraphics[width=0.45\textwidth,viewport=30 30 700 600,clip]{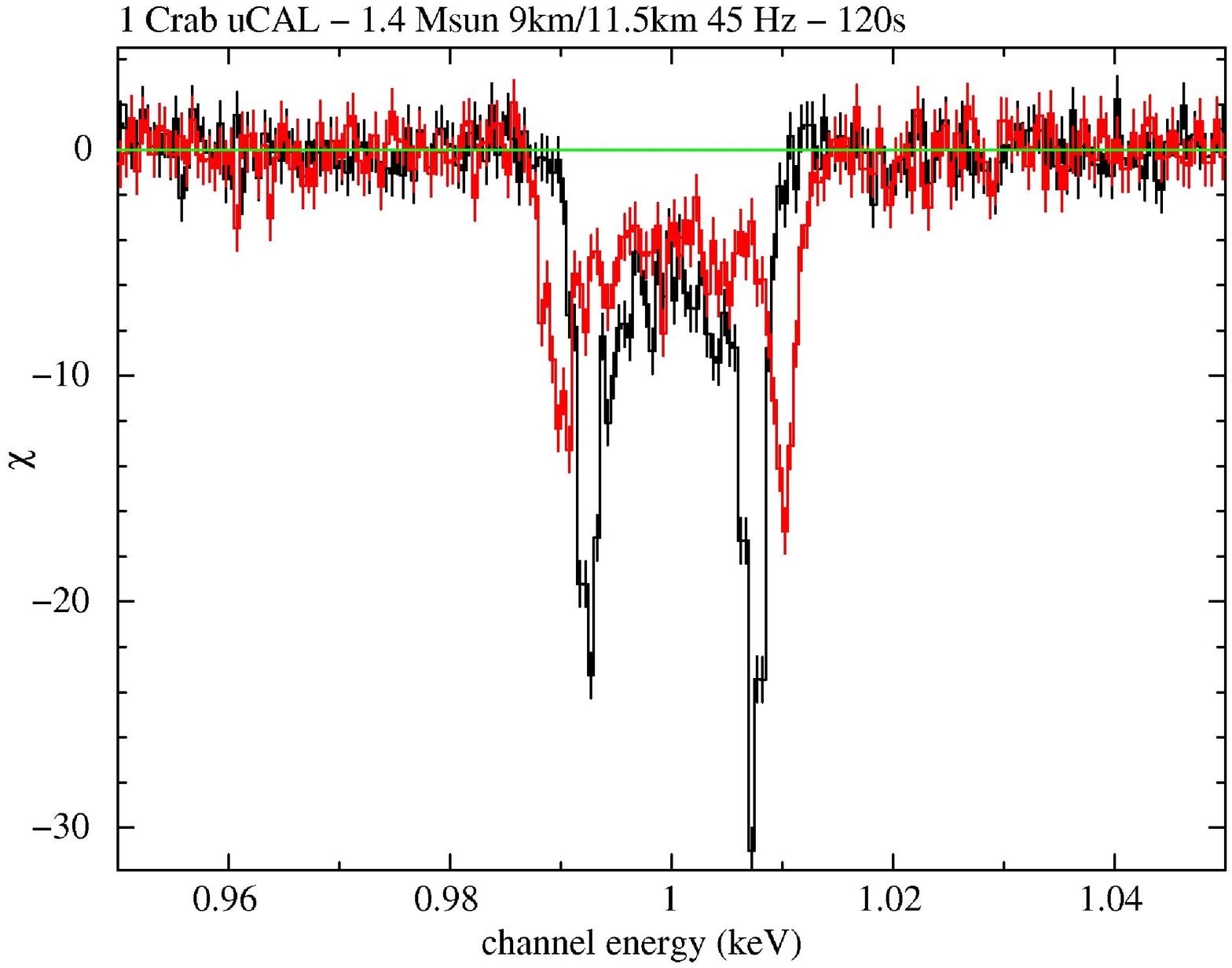}
\hfill
\includegraphics[width=0.45\textwidth,viewport=30 30 700 600,clip]{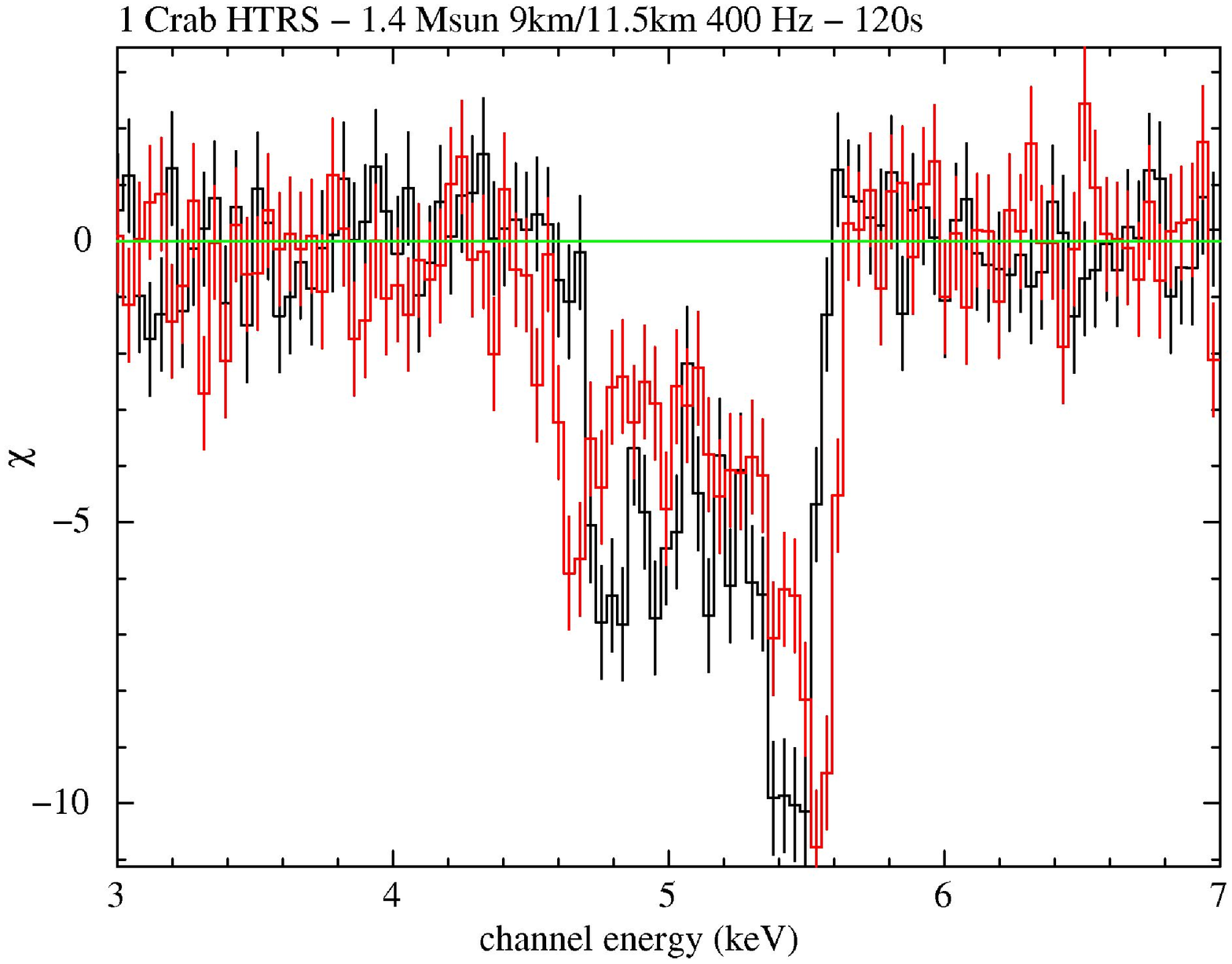}\\
\caption{\small{
{\bf High resolution X-ray spectroscopy of the photospheric emission of a hot neutron star is sensitive to the fundamental stellar parameters}, through the effects of pressure broadening, relativistic kinematics (rotation, Doppler shift, time dilation, beaming), and general relativity (light bending around the star, gravitational redshift, frame dragging) on atomic absorption lines (\"Ozel and Psaltis 2003; Bhattacharyya, Miller, and Lamb 2006). The absorption line spectrum of a $1.4 M_{\odot}$ neutron star, spinning at 45 Hz, showing the effects of rotational Doppler-splitting, observed at 2 eV spectral resolution ({\it left panel}), in 120 sec of exposure of a moderately bright X-ray burst with the microcalorimeter spectrometer as envisioned for the {\it IXO} mission. Black and red histograms refer to a star with a radius of 9 and 11.5 km, respectively. Emission is concentrated in a hot equatorial belt, seen at 5 degree inclination. The absorption line is Fe XXVI H$\alpha$. {\bf High time resolution spectroscopy can phase-resolve the Doppler broadening} of a rapidly spinning star (400 Hz) if the surface emission is azimuthally asymmetric. With $\sim 100$ eV energy resolution, and sub-msec time resolution (such as for the fast timing instrument on {\it IXO}), the Doppler profiles {\it in the right hand panel} will be phase-resolved, allowing unambiguous determination of the line broadening mechanism, and an absolute radius measurement (Fe XXVI Ly$\alpha$; same stellar parameters as before).
}
}
\label{}
\end{figure}

The ability to time- and energy-resolve emission from bursting or spinning stars provides unique observational leverage. In particular, a capability to perform rapid (tens of microseconds or less) spectroscopy allows resolving (and uniquely identifying) the severe Doppler broadening associated with high stellar spin frequencies known to occur in most LMXBs. 
For objects with a known spin period, the Doppler broadening provides a direct measurement of the stellar radius. GR light bending effects on the phase modulation (of the absorption lines as well as the total flux) again produce a measurement of the acceleration of gravity at the surface. In cases where an atomic absorption line is detected, a redshift measurement is sufficient (with redundancy in the pressure broadening). Note that the magnetic field strengths in LMXB's are small enough 
($< 10^9$ G) that Zeeman splitting is not important.

High-speed, high-time resolution photometry of quasi-periodic intensity fluctuations associated with the inner accretion disk in accreting neutron stars can yield absolute sizes, if light-travel time effects can be resolved (reverberation). 
A time-delay spectrum within the
frequency range of the variability immediately provides the
physical size of the inner rim of the accretion disk, where
reprocessed soft photons come from, and hence an upper limit to the
neutron-star radius in km, independent of the other stellar parameters
(Gilfanov et al. 2003; Vaughan et al. 1997, 1998). Likewise, Fe K line emission reverberation
yields the same information (Cackett et al. 2008).

There is multiple redundancy in several techniques, and we will have a choice Fof techniques to cover a range of expected stellar spin periods (spin frequencies up to several hundred Hz have been measured in LMXBs). 
Most of the necessary theoretical development (full radiative transfer neutron star atmosphere models,
effects of light bending on distant observer flux spectrum, etc.) is in place.

The precision required to make definitive measurements of the neutron star mass-radius relation is within reach with currently feasible technology. The energy resolution of cryogenic X-ray spectrometers, such as microcalorimeters, is sufficient to detect photospheric absorption lines and measure their profiles. The count rate capability, time resolution, and CCD-style energy resolution of Si drift detectors meet the requirements of fast energy-resolved timing. The {\it International X-ray Observatory}'s capabilities, with a high energy resolution microcalorimeter spectrometer\footnote{the X-ray Microcalorimeter Spectrometer, or XMS instrument}, and a high time-resolution medium energy-resolution spectrometer\footnote{the High Time Resolution Spectrometer, or HTRS instrument} are ideal for a definitive 
solution to the Cold Equation of State problem. With the effective area, energy resolution, and timing capability of {\it IXO}, the list of potential targets is at least a dozen deep for X-ray burst sources, and several quiescent LMXBs will be observable as well.

Finally, the X-ray techniques will complement possible results from radio pulsar observations. The binary radio pulsar PSR J$0737-3039$ is known to exhibit a pulse arrival time evolution of one of the two members that may signal relativistic spin-orbit coupling of the binary. If that is indeed correct, the moment of inertia of this neutron star may be measurable, in addition to its mass, and these two quantities together constrain the equation of state (Lyne et al. 2004; Kramer and Stairs 2008; see also Lattimer and Prakash 2007). The mass of this neutron star is close to the average mass of observed for radio pulsar neutron stars, and so this measurement may have limited leverage on the problem of distinguishing between hadronic and strange equations of state (see Figure 2). Neutron stars in mass-transferring binaries will give us access to a wider range of neutron star masses (of order a solar mass of material can be transferred over the lifetime of an LMXB), to address this fundamental issue. 

\section{Additional Science}

We now know that neutron stars with surface magnetic field strengths in excess of $10^{14}$ Gauss exist (Duncan and Thompson 1992). At field strengths exceeding $m_e^2 c^3/\hbar e = 4.4 \times 10^{13}$ G, QED predicts novel effects, such as vacuum birefringence; in the presence of matter, resonant polarization mode conversion will occur (Lai and Ho 2003, and references therein). Direct X-ray spectroscopy of such objects may reveal proton cyclotron resonance absorption at photon energy $0.6 (1+z)^{-1}(B/10^{14}\ {\rm G})$ keV (Bezchastnov et al. 1996), in spite of the relatively small transition probability ({\it e.g.} Ho and Lai 2001). The shape and angular dependence of the photospheric X-ray spectrum will reflect these various effects ({\it e.g.} van Adelsberg and Lai 2006). Alternatively, the spectrum could show atomic absorption features ({\it e.g.} Zavlin and Pavlov 2002; Hailey and Mori 2002). Moreover, the surface emission of all strongly magnetized neutron stars should exhibit a strong, energy-dependent polarization (Pavlov and Zavlin 2000). With polarizations of up to tens of percent, and a potentially dramatic phase dependence, X-ray polarimetry can probe this regime of QED for the first time.

Finally, should we find neutron stars with a combination of mass and radius that is in conflict with all
physically plausible equations of state (for instance, a $1.4M_{\odot}$, 18 km object; see Figure 2), we would have evidence that we are probing modifications to the equation of hydrostatic equilibrium, and not the equation of state: the neutron star mass-radius relation measurements can reveal deviations from the standard theory of gravity. The measurement technique is equally sensitive to neutron star parameters both inside and outside the region of the mass-radius plane covered by the equation of state margins of uncertainty, and so is also an entirely novel probe of gravity itself (Psaltis 2008).
\vskip 0.3 in
{\bf References}\\
\scriptsize{
Bezchastnov, V. G., Pavlov, G. G., and Shibanov, Yu. A., 1996, in {\it Gamma Ray Bursts}, 

AIP Conf. Proc., {\bf 384}, 907\\
Bhattacharyya, S., Miller, M. C., and Lamb, F. K., 2006, {\it Astrophys. J.}, {\bf 664}, 1085\\
Blandford, R. D., 1993, in {\it Pulsars as Physics
Laboratories}, Oxford UP, 1993\\
Cackett, E. M., et al. 2008, {\it Astrophys. J.}, {\bf 674}, 415\\
Cottam, J., Paerels, F., and M\'endez, M., 2002, {\it Nature}, {\bf 420}, 51\\
Duncan, R. C., and Thompson, C., 1992,  {\it Astrophys. J. (Letters)}, {\bf 392}, L9\\
D\"urr, S., et al., 2008, {\it Science}, {\bf 322}, 1224\\
Gilfanov, M., Revnivtsev, M., and Molkov, S.,  2003, {\it Astron. Astrophys.}, {\bf 410}, 217\\
Guillot, S., Rutledge, R. E., Bildsten, L., Brown, E. F., Pavlov, G. G., 

and Zavlin, V. E., 2009, {\it Mon. Not. Royal Astr. Soc.}, {\bf 392}, 665\\
Hessels, J. W. T., et al., 2006, {\it Science}, {\bf 311}, 1901\\
Hailey, C., and Mori, K., 2002,  {\it Astrophys. J. (Letters)}, {\bf 578}, L133\\
Ho, W. C. G., and Lai, D., 2001, {\it Mon. Not. Royal Astr. Soc.}, {\bf 327}, 1081\\
Kaaret, P., et al., 2006, {\it Astrophys. J. (Letters)}, {\bf 257}, L97\\
Kaplan, D. L., van Kerkwijk, M. H., and Anderson, J. 2002, {\it Astrophys. J.}, {\bf 571}, 447\\
-----, 2007, {\it Astrophys. J.}, {\bf 660}, 1428\\
Kaspi, V. M., Roberts, M. S. E., and Harding, A. K., 2006, in {\it Compact Stellar X-ray Sources}, 

W. H. G. Lewin and M. van der Klis (Eds.), Cambridge UP, 2006 (astro-ph/0402136)\\
Kramer, M., and Stairs, I. H., 2008, {\it Ann. Rev. Astron. Astrophys.}, {\bf 46}, 541\\
Lai, D., and Ho, W. C. G., 2003, {\it Phys. Rev. Letters}, {\bf 91}, 071101\\
Landau, L. D., 1932, {\it Physikalische Zeitschrift der Sowjetunion}, {\bf 1}, 285\\
Landau, L. D., 1938, {\it Nature}, {\bf 141}, 333\\
Lattimer, J. M., and Prakash, M., 2007, {\it Physics Reports}, {\bf 442}, 109\\
Lyne, A. G., et al., 2004, {\it Science}, {\bf 303}, 1153\\
\"Ozel, F., and Psaltis, D., 2003,  {\it Astrophys. J.}, {\bf 529}, 1011\\
Pavlov, G. G., and Zavlin, V. E., 2000, {\it Astrophys. J. (Letters)}, {\bf 582}, L31\\
Pavlov, G. G., Kargaltsev, O., Wong, J. A., and Garmire, G. P., 2008, astro-ph/0803.0761\\
Psaltis, D., 2008, {\it Living Reviews in Relativity}, {\bf 11}, no. 9 (cited on 1-21-2009)\\
Strohmayer, T., and Bildsten, L., 2006, in {\it Compact Stellar X-ray Sources}, 

W. H. G. Lewin and M. van der Klis (Eds.), Cambridge UP, 2006  (astro-ph/0301544)\\
Tannenbaum, M. J., 2006, {\it Rep. Prog. Phys.}, {\bf 69}, 2005\\
van Adelsberg, M., and Lai, D., 2006,  {\it Mon. Not. Royal Astr. Soc.}, {\bf 373}, 1495\\
Vaughan, B. A., et al., 1997, {\it Astrophys. J. (Letters)}, {\bf 483}, L115; 
erratum, 1998, {\it Astrophys. J. (Letters)}, {\bf 509}, L145\\
Walter, F. M., and Lattimer, J. M., 2002,  {\it Astrophys. J. (Letters)}, {\bf 576}, L145\\
Weisberg, J. M., and Taylor, J. H., 2005, in {\it Binary Radio Pulsars}, ASP Conf. Proc., {\bf 328}, 25\\
Zavlin, V. E., and Pavlov, G. G., 2002,  
MPE Reports, {\bf 278}, 273 (astro-ph/0206025)
}


\end{document}